\documentclass[aps,prl,twocolumn,footinbib,showpacs,floatfix]{revtex4}
\usepackage{amssymb}
\usepackage[pdftex]{graphicx}	
	\DeclareGraphicsExtensions{.pdf,.png,.jpg}
\usepackage[applemac]{inputenc}
\usepackage{amsmath,amssymb,amsthm}
\usepackage{mathbbol,bbm}
\usepackage{graphicx,float}
\usepackage[final]{showkeys}
\usepackage{bm,dsfont}

\def\ket#1{\left|#1\right>}
\def\bra#1{\left<#1\right|}
\def\Tr{ {\rm{Tr }}}

\usepackage{booktabs}

\begin{document}
\title{Uhlmann Phase as a Topological Measure for One-Dimensional Fermion Systems}
\author{O. Viyuela, A. Rivas and M.A. Martin-Delgado}
\affiliation{Departamento de F\'{\i}sica Te\'orica I, Universidad Complutense, 28040 Madrid, Spain}

\vspace{-3.5cm}

\begin{abstract}
We introduce the Uhlmann geometric phase as a tool to characterize symmetry-protected topological phases in 1D fermion systems, such as topological insulators and superconductors. Since this phase is formulated for general mixed quantum states, it provides a way to extend topological properties to finite temperature situations. We illustrate these ideas with some  paradigmatic models and find that there exists a critical temperature $T_c$ at which the Uhlmann phase goes discontinuously and abruptly to zero. This stands as a borderline between two different topological phases as a function of the temperature. Furthermore, at small temperatures we recover the usual notion of topological phase in fermion systems.
\end{abstract}


\maketitle


\noindent {\it 1.Introduction.---} Geometric phases have played an essential role in many quantum phenomena
since its modern discovery by Berry \cite{Berry84} (see also Refs. \cite{Simon85,WilczekBook}). An emblematic example is the characterization of the transversal conductivity $\sigma_{xy}$ in the quantum Hall effect by means of the integral of the Berry curvature over the two-dimensional
Brillouin zone (BZ), in units of $\frac{e^2}{h}$. This is the celebrated TKNN formula \cite{TKNN} that has become a key ingredient
in the characterization in the newly emerging field of topological insulators \cite{rmp1,rmp2}.
Recently, the experimental measurement of a Berry phase in a one-dimensional optical lattice (Zak phase \cite{Zak89}) simulating the different
phases of polyacetylene \cite{Atala_et_al_12} has opened the way to extend the applications of geometric phases to study topological properties beyond the realm of condensed-matter systems.

A fundamental problem in the theory and applications of geometrical phases is its extension from pure quantum states (Berry) to mixed quantum states described by density matrices. Uhlmann was first to mathematically address this issue \cite{Uhlmann86} and to provide a satisfactory solution \cite{Uhlmann89, Uhlmann91, Hubner93, Uhlmann96}. For more than a decade, there has been a renewed interest in studying geometric phases for mixed states and under dissipative evolutions from the point of view of quantum information \cite{Soqvist_et_al_00}, and more inequivalent definitions have been introduced \cite{Bhandari02, Anandan_et_al02, Slater02}. This has culminated with the first experimental measurement of a geometric phase for mixed quantum states of one system qubit and one ancillary qubit  with NMR techniques \cite{Du_et_al03}.

In addition, the role played by external dissipative effects and thermal baths in topological insulators and superconductors has attracted much interest both in quantum simulations with different platforms and in condensed matter \cite{Muller_et_al12,Bardyn_et_al13, Bardyn_et_al12, Kraus_et_al12, Mazza_et_al12, Viyuela_et_al12, Rivasl_et_al13, Garate13, 1D_TI_OL_12, 1D_TI_QC_11, Tarruell_et_al12}. In this work we show that the Uhlmann geometric phase is endowed with a topological structure when applied to one-dimensional fermion systems. More concretely,

\noindent i/  We show that the Uhlmann phase allows us to characterize topological insulators and superconductors at both zero and finite temperature.

\noindent ii/ We find a finite critical temperature $T_c$ below which the Uhlmann phase is constant and nonvanishing. At $T_c$ there is a discontinuity, and above it the topological behavior ceases to exist. This kind of behavior is very relevant and not present in other formulations.

\noindent iii/ We study 1D paradigmatic models such as the Creutz ladder (CL) \cite{Creutz99, Bermudez_et_al09}, the Majorana chain (MC) \cite{Kitaev01} and the polyacetylene (SSH) \cite{SSH, Rice_Mele82}. A summary of the basic results of this paper is presented in Table I. Notably, at the limit of zero temperature the Uhlmann phase recovers the usual notion of topological order as given by the Berry phase. Moreover, when the three models are in a flat-band regime the critical temperature is universal, Eq. \eqref{Tc}.

\begin{figure}[t]
	\includegraphics[width=\columnwidth]{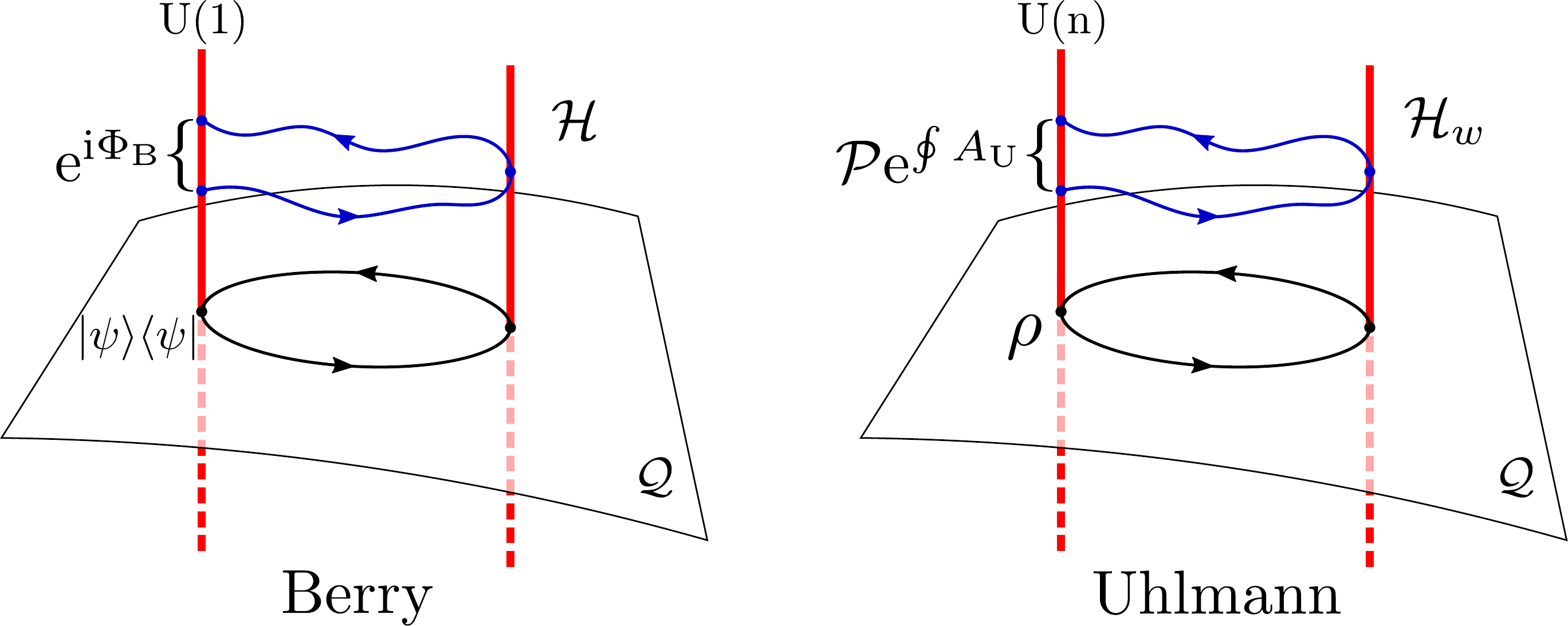}
	\caption{Comparison of the Berry and Uhlmann approaches. The usual U(1)-gauge freedom is generalized to the U(n)-gauge freedom of the amplitudes in the Uhlmann's approach. Thus, according to Berry, after a closed loop in the set $\mathcal{Q}$, a pure state carries a simple phase factor $\Phi_{\rm B}$. However, for mixed states the amplitude carries a unitary matrix $\mathcal{P}{\rm e}^{\oint A_{\rm U}}$.}
	\label{Fibrados}
\end{figure}

The Uhlmann approach is based on the concept of amplitude. An amplitude for some density matrix $\rho$ is any of the matrices $w$ such that
\begin{equation}\label{ww^daga}
\rho=ww^\dagger.
\end{equation}
The key idea behind this definition is that the amplitudes form a Hilbert space $\mathcal{H}_w$ with the Hilbert-Schmidt product $(w_1,w_2):=\Tr(w_1^\dagger w_2)$. On the contrary, the set of density matrices $\mathcal{Q}$ is not a linear space. From Eq. \eqref{ww^daga}, we see that there is a ${\rm U}(n)$-gauge freedom in the choice of the amplitude ($n$ is the dimension of the space):  $w$ and $wU$  are amplitudes of the same state for some unitary operator $U$. Note the parallelism with the usual ${\rm U}(1)$-gauge freedom of pure states, where $|\psi\rangle$ and ${\rm e}^{{\rm i}\phi}|\psi\rangle$ represent the same physical state, i.e., the same density matrix given by $|\psi\rangle\langle\psi|$. Thus the usual gauge freedom can be seen as a particular case of the amplitude ${\rm U}(n)$-gauge freedom.

\begin{table}[t]
\begin{tabular}{|c|c|c|c|}
\toprule
\hline
\hline
\multicolumn{4}{c}{Topological Measures in 1D Fermion Models} \\
\hline
\cmidrule(r){1-2}
& \ \ CL \ \ & \ \ MC \ \ & \ SSH \ \\
\hline
\midrule
\ \ Winding Number $(T=0)$ \ \ & 1 & 1 & 1 \\
\hline
Berry Phase $(T=0)$ & $\pi$ & $\pi$ & $\pi$ \\
\hline
Uhlmann Phase $(T< T_c)$ & $\pi$ & $\pi$ & $\pi$ \\
\hline
Uhlmann Phase $(T> T_c)$ & 0 & 0 & 0 \\
\hline
\hline
\bottomrule
\end{tabular}
\caption{Comparison of Hamiltonian winding number, Berry and Uhlmann phases for non-trivial topological regimes in the Creutz ladder (CL), Majorana chain (MC) and polyacetilene (SSH) 1D fermion models.}
\end{table}

An amplitude is nothing but another way to see the concept of purification. Indeed, by the polar decomposition theorem, we parametrize the possible amplitudes of some density matrix $\rho$ as $w=\sqrt{\rho}U$. Because of the spectral theorem, $\rho=\sum_jp_j|\psi_j\rangle\langle\psi_j|$, we have $w=\sum_j\sqrt{p_j}|\psi_j\rangle\langle\psi_j|U$. Let us define the following isomorphism between the spaces $\mathcal{H}_w$ and $\mathcal{H}\otimes\mathcal{H}$: $w=\sum_j\sqrt{p_j}|\psi_j\rangle\langle\psi_j|U\longleftrightarrow |w\rangle=\sum_j\sqrt{p_j}|\psi_j\rangle \otimes U^{\rm t}|\psi_j\rangle$ (here the transposition is taken with respect to the eigenbasis of $\rho$). The property $\rho=ww^\dagger$ is now written as
\begin{equation}\label{purification}
\rho=\Tr_2(|w\rangle\langle w|).
\end{equation}
Here, $\Tr_2$ denotes the partial trace over the second Hilbert space of $\mathcal{H}\otimes\mathcal{H}$. In other words, any amplitude $w$ of some density matrix $\rho$ can be seen as a pure state $|w\rangle$ of the enlarged space $\mathcal{H}\otimes\mathcal{H}$, with partial trace equal to $\rho$. Thus, $|w\rangle$ is a purification of $\rho$.

Let us consider a family of pure states $|\psi_{\bm{k}}\rangle\langle\psi_{\bm{k}}|$ and some trajectory in parameter space, $\{\bm{k}(t)\}_{t=0}^1$, such that the initial and final states are the same. This induces a trajectory on the Hilbert space $\mathcal{H}$, $|\psi_{\bm{k}(t)}\rangle$, and since the path on $\mathcal{Q}$ is closed, the initial and final vectors are equivalent up to some $\Phi$, $|\psi_{\bm{k}(1)}\rangle={\rm e}^{{\rm i} \Phi} |\psi_{\bm{k}(0)}\rangle$. Provided the transportation of the vectors in $\mathcal{H}$ is done following the Berry parallel transport condition (i.e. no dynamical phase is accumulated) $\Phi$ is the well-known Berry phase $\Phi_{\rm B}$. This depends only on the geometry of the path and can be written as $\Phi_{\rm B}=\oint A_{\rm B}$, where $A_{\rm B}:={\rm i}\sum_{\mu}\langle\psi_{\bm{k}}|\partial_\mu\psi_{\bm{k}}\rangle dk_\mu$ is the Berry connection form $(\partial_\mu :=\partial/\partial k_\mu)$. Similarly, we may have a closed trajectory of not necessarily pure density matrices $\rho_{\bm{k}}$, which in turn induces a trajectory on the Hilbert space $\mathcal{H}_w$, $w_{\bm{k}(t)}$. Again, since the path on $\mathcal{Q}$ is closed, the initial and final amplitudes must differ just in some unitary transformation $V$, $w_{\bm{k}(1)}=w_{\bm{k}(0)}V$. Hence, by analogy to the pure state case, Uhlmann defines a parallel transport condition such that $V$ is given by $V=\mathcal{P}{\rm e}^{\oint A_{\rm U}}U_0$; where $\mathcal{P}$ stands for the path ordering operator, $A_{\rm U}$ is the Uhlmann connection form and $U_0$ is the gauge taken at $\bm{k}(0)$. We have illustrated this parallelism between the Berry and Uhlmann approach in Fig. \ref{Fibrados}.

The Uhlmann parallel transport condition asserts that for some point $\rho_{\bm{k}(t)}$ with amplitude $w_{\bm{k}(t)}$, the amplitude $w_{\bm{k}(t+dt)}$ of the next point in the trajectory, $\rho_{\bm{k}(t+dt)}$, is the closest \cite{footnote1} to $w_{\bm{k}(t)}$ among the possible amplitudes of $\rho_{\bm{k}(t+dt)}$. With this rule, it is possible to obtain some explicit formulas for $A_{\rm U}$. Concretely, in the spectral basis of $\rho_{\bm{k}}=\sum_jp^j_{\bm{k}}|\psi^j_{\bm{k}}\rangle\langle\psi^j_{\bm{k}}|$, one obtains \cite{Hubner93}
\begin{equation}\label{dAfinal}
A_{\rm U}=\sum_{\mu,i,j}|\psi^i_{\bm{k}}\rangle\frac{\langle\psi^i_{\bm{k}}|\left[(\partial_\mu\sqrt{\rho_{\bm{k}}}),\sqrt{\rho_{\bm{k}}}\right]|\psi^j_{\bm{k}}\rangle}{p^i_{\bm{k}}+p^j_{\bm{k}}}\langle\psi^j_{\bm{k}}|dk_\mu.
\end{equation}
Note that this connection form has only zeroes on its diagonal and is skew adjoint, so that the Uhlmann connection is special unitary.
The Uhlmann geometric phase along a closed trajectory $\{\bm{k}(t)\}_{t=0}^1$ is defined as
\begin{equation}
\Phi_{\rm U}:=\arg\langle w_{\bm{k}(0)}|w_{\bm{k}(1)}\rangle
=\arg \Tr\left[w^\dagger_{\bm{k}(0)}w_{\bm{k}(1)}\right].
\end{equation}
By the polar decomposition theorem, we may write $w_{\bm{k}(0)}=\sqrt{\rho_{\bm{k}(0)}}U_0$, $w_{\bm{k}(1)}=\sqrt{\rho_{\bm{k}(0)}}V$, so that
\begin{equation}
\Phi_{\rm U}=\arg \Tr\left[\rho_{\bm{k}(0)} \mathcal{P}{\rm e}^{\oint A_{\rm U}}\right].
\label{phiU}
\end{equation}

As aforementioned, in this work we shall focus on the Uhlmann phase in 1D fermion models. For such systems, $\bm{k}\equiv k$ is the one-dimensional crystalline momentum living in a $S^1$-circle BZ. Thus, because of the non-trivial topology of $S^1$, geometric phases after a loop in $k$ acquire a topological sense.

\noindent {\it 2. Fermionic Systems and Uhlmann Phase.---}
Consider two-band Hamiltonians within the spinor representation $\Psi_k=(\hat{a}_k,\hat{b}_k)^{\rm t}$, where $\hat{a}_k$ and $\hat{b}_k$ stands for two species of fermionic operators.
For superconductors the spinor $\Psi_k$ is constructed out of a Nambu transformation of paired fermions with opposite crystalline momentum \cite{ASBook}. The Hamiltonian is a quadratic form $H=\sum_k\Psi_k^{\dagger}H_k\Psi_k$ and $H_k$ is a $2\times2$ matrix:
\begin{equation}\label{Hk}
H_k=f(k)\mathds{1} +\frac{\Delta_k}{2}{\bm n}_k\cdot{\bm \sigma}.
\end{equation}
Here, $\bm{\sigma}=(\sigma_x,\sigma_y,\sigma_z)$ are the Pauli matrices, $\Delta_k$ corresponds to the gap of $H_k$ and $f(k)$ denotes some function of $k$. The unit vector ${\bm n}_k=(\sin{\theta}\cos{\phi},\sin{\theta}\sin{\phi},\cos{\theta})$ is called the `winding vector" where $\theta$ and $\phi$ are $k-$dependent spherical coordinates.
The band eigenvectors of $H_k$ can be written as
\begin{equation}
\ket{u_{-}^k}=\begin{pmatrix}-\text{e}^{-{\rm i}\phi(k)}\sin{\frac{\theta(k)}{2}}\\ \cos{\frac{\theta(k)}{2}} \end{pmatrix},~~\ket{u_{+}^k}=\begin{pmatrix}\text{e}^{-{\rm i}\phi(k)}\cos{\frac{\theta(k)}{2}}\\ \sin{\frac{\theta(k)}{2}} \end{pmatrix},
\label{us}
\end{equation}

\begin{figure*}[t]
	\includegraphics[width=\textwidth]{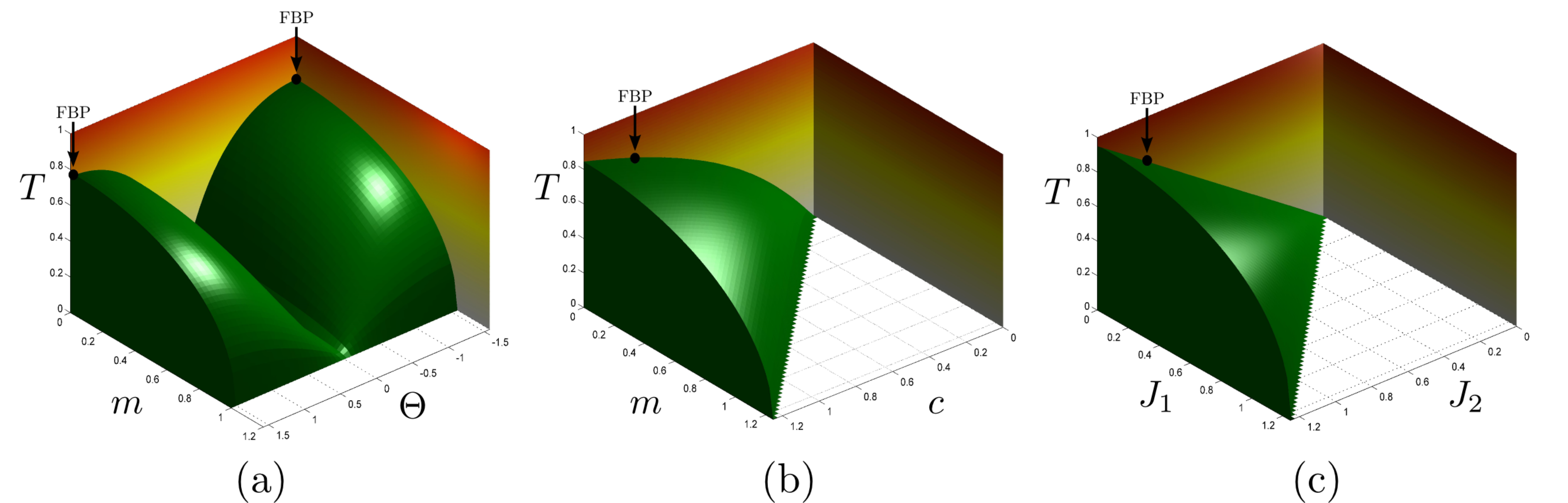}
	\caption{Uhlmann topological phases for the Creutz Ladder (CL) (a),  Majorana Chain (MC) (b) and Polyacetilene (SSH) (c). They are $\pi$ inside the green volume and zero outside.
	The flat-band points  FBP are indicated with an arrow and are universal. Natural units have been taken. In addition, for the CL and the MC we have fixed the horizontal hopping $2R=1$ and the superconducting pairing $|M|=1$ respectively.}
	\label{fig_U}
\end{figure*}
\vspace{0.1cm}

If the thermalization process preserves particle number, and the Fermi energy is set in the middle of the gap, the equilibrium (thermal) state is given by $\rho_{\beta}=\prod_k\rho_k^\beta$, with
\begin{equation}
\rho_k^\beta=\frac{{\rm e}^{-H_k/T}}{\Tr\left({\rm e}^{-H_k/T}\right)}=\frac{1}{2}\left[\mathds{1}-\tanh{\left(\frac{\Delta_k}{2T}\right)}\hat{{\bm n}}_k\cdot{\bm \sigma}\right],
\label{steady2}
\end{equation}
where $T=1/\beta$ denotes temperature.

By the use of Eq. \eqref{us}, the Uhlmann connection \eqref{dAfinal} for $\rho_k^\beta$ turns out to be
\begin{equation}
A^{k}_{\rm U}=m^k_{12}\langle u_{-}^k\ket{\partial_k u_{+}^k}\ket{u_{-}^k}\bra{u_{+}^k}dk+\text{h.c.}
\label{Arho2}
\end{equation}
where $m^k_{12}:=1-{\rm sech}\left(\frac{\Delta_k}{2T}\right)$.

Besides, it is well known that discrete symmetries represent a way to classify topological insulators and superconductors \cite{Ludwig,Kitaev_2009}. Furthermore, for the models considered throughout this paper, symmetries impose a restriction on the movement of ${\bm n}_k$ to some plane as a function of $k$, making only two of its components ${n}^i_k$ and ${n}^j_k$ with $i\not=j$, different from zero. Therefore, we have a nontrivial mapping $\mathit{S}^1\longrightarrow \mathit{S}^1$, characterized by a winding number $\omega_1$. This is defined using the angle $\alpha$ covered by ${\bm n}_k$ when it winds around the unit circle $\mathit{S}^1$, and takes the form
\begin{equation}
\omega_1:=\frac{1}{2\pi}\oint d\alpha=\frac{1}{2\pi}\oint\bigg(\frac{\partial_k {n}_k^i}{n_k^j}\bigg)dk,
\label{w1}
\end{equation}
where we have used that $\alpha:=\arctan\big({n}^i_k/{n}^j_k\big)$.

Moreover, using Eqs. \eqref{us} and \eqref{w1} with \eqref{Arho2}, and simplifying Eq. \eqref{phiU} we obtain an expression for the Uhlmann phase in terms of $\omega_1$, the temperature, and parameters of the Hamiltonian
\begin{equation}
\Phi_{\rm U}=\arg\Bigg\{\cos(\pi\omega_1)\cos\bigg[\oint\bigg(\frac{\partial_k {n}_k^i}{2n_k^j}\bigg)\text{sech}\left(\frac{\Delta_k}{2T}\right)dk\bigg]\Bigg\}.
\label{phiU2}
\end{equation}
Particularly, in the limit $T\rightarrow0$,
\begin{equation}
\Phi_{\rm U}^{0}=\arg[\cos(\pi \omega_1)].
\label{phiUb3}
\end{equation}
Note that for the trivial case $\omega_1=0$, the Uhlmann phase is zero as well. However, for nontrivial topological regions $\omega_1=\pm1$, we obtain $\Phi_{\rm U}^0=\pi$. Thus, the topological order as accounted by $\Phi_{\rm U}^{0}$ coincides to the standard notion measured by $\omega_1$. In the following, we compute $\Phi_{\rm U}$ at finite temperature for the three aforementioned models of topological insulators and superconductors.

\noindent {\it 3. Creutz Ladder.---}
This model \cite{Creutz99} is representative for a topological insulator \cite{Bermudez_et_al09,Viyuela_et_al12} with AIII symmetry \cite{Ludwig,Kitaev_2009}. It describes the dynamics of spinless electrons moving in a ladder as dictated by the following Hamiltonian:
\begin{eqnarray}
H_{\rm CL}&=&-\sum_{n=1}^L[R(\text{e}^{-{\rm i}\Theta}a^{\dagger}_{n+1}a_n+\text{e}^{{\rm i}\Theta}b^{\dagger}_{n+1}b_n)+\nonumber\\
&+&R(b^{\dagger}_{n+1}a_n+a^{\dagger}_{n+1}b_n)+Ma^{\dagger}_{n}b_n+{\rm h.c.}],
\label{HCL}
\end{eqnarray}
where $a_n$ and $b_n$ are fermionic operators associated to the $n$-th site of an upper and lower chain, respectively. The hopping along horizontal and diagonal links is given by $R>0$, and the vertical one by $M>0$. In addition, a magnetic flux $\Theta\in[-\pi/2,\pi/2]$ is induced by a perpendicular magnetic field. For nonzero magnetic flux $\Theta\not=0$ and small vertical hopping, $m:=\frac{M}{2R}<1$, the system has localized edge states at the two ends of the open ladder \cite{Creutz99}. Interestingly, there exists an experimental proposal for this model with optical lattices  \cite{CL_OL_12}.

In momentum space, $H_{\rm CL}$ can be written in the form of Eq. \eqref{Hk} with (in units of $2R=1$)
\begin{eqnarray}
{\bm n}_k&=&\frac{2}{\Delta_k}(m+\cos{k},0,\sin{\Theta}\sin{k}),\nonumber\\
\Delta_k&=&2\sqrt{(m+\cos{k})^2+\sin^2{\Theta}\sin^2{k}},
\label{parCL}
\end{eqnarray}
which in the spinor decomposition made in \eqref{Hk} implies $\phi=0,\pi$.

By the means of \eqref{phiU2} we compute the value of the Uhlmann phase (which can only be equal to $\pi$ or 0) as function of parameters $\Theta$, $m$ and the temperature $T$ [see Fig. \ref{fig_U}(a)]. At $T\rightarrow0$ the topological region coincides with the usual topological phase, $\Phi^0_{\rm U}=\Phi_{\rm B}=\pi$ for $m\in[0,1]$ and $\Theta\in[-\frac{\pi}{2},\frac{\pi}{2}]$, as expected. However there exists a critical temperature $T_c$ for any value of the parameters at which the system is not topological in the Uhlmann sense anymore and $\Phi_{\rm U}$ goes abruptly to zero. The physical meaning of this $T_c$ relies on the existence of some critical momentum $k_c$ splitting the holonomy into two disequivalent topological components according to the value taken by $k$ when performing the closed loop, $\Phi_{\rm U}(k<k_c)=0$ and $\Phi_{\rm U}(k>k_c)=\pi$ respectively. In the trivial topological regime there is only one component with $\Phi_{\rm U}=0$ for every point along the trajectory. Thus, this structure of the Uhlmann amplitudes accounts for a topological kink \cite{Coleman} in the holonomy along the BZ. Further details about the presence or absence with temperature of this topological kink can be seen in the Supplementary Material \cite{SM}.

Interestingly, at $m=0$ and $\Theta=\pm\frac{\pi}{2}$ (see the arrows in Fig. \ref{fig_U}), the edge states become completely decoupled from the system dynamics. When considering periodic boundary conditions, this translates into having \emph{flat bands} in the spectrum. For these flat-band points (FBPs) the critical temperature $T_c$ only depends on the constant value of the gap $\Delta_k=2$ and can be analytically computed. The result is the same for the three models analyzed in this work,
\begin{equation}
\label{Tc}
T_c=\frac{1}{\ln{(2+\sqrt{3})}},
\end{equation}
which is approximately $38\%$ of the gap.

\noindent {\it 4. Majorana Chain.---}
Consider a model of spinless fermions with {\it p}-wave superconducting pairing, hopping on a $L$-site one-dimensional chain. The Hamiltonian of this system introduced by Kitaev \cite{Kitaev01} is
\begin{equation}
H_{\rm MC}=\sum^L_{j=1}\left(-Ja^{\dagger}_ja_{j+1}+M a_ja_{j+1}-\textstyle{\frac{\mu}{2}}a_j^{\dagger}a_j+\text{H.c.}\right),
\label{Hkit}
\end{equation}
where $\mu>0$ is the chemical potential, $J>0$ is the hopping amplitude, the absolute value of $M=|M|\text{e}^{{\rm i}\Theta}$ stands for the superconducting gap, and $a_j$ $(a_j^{\dagger})$ are annihilation (creation) fermionic operators.

For convenience, we may redefine new parameters $m:=\frac{\mu}{2|M|}$ and $c:=\frac{J}{|M|}$, and take $\Theta=0$. It can be shown \cite{Kitaev01} that the system has nonlocal Majorana modes at the two ends on the chain if $m<c$, which corresponds to nonvanishing $\omega_1$ and $\Phi_{\rm B}$ when taking periodic boundary conditions. Thus, in momentum space, $H_{\rm MC}$ can be written in the form of Eq. \eqref{Hk} using the so-called Nambu spinors $\Psi_k=\big(a_k,a^{\dagger}_{-k}\big)^{\rm t}$:
\begin{eqnarray}
{\bm n}_k&=&\frac{2}{\Delta_k}(0,-\sin{k},-m+c\cos{k}),\nonumber\\		
\Delta_k&=&2\sqrt{(-m+c\cos{k})^2+\sin^2{k}},
\label{parMC}
\end{eqnarray}
in units of $|M|=1$. This in \eqref{Hk} implies $\phi=\pm\frac{\pi}{2}$.

In analogy to the CL case, we calculate the Uhlmann phase as a function of parameters $m$, $c$ and the temperature $T$ [see Fig. \ref{fig_U}(b)]. On the one hand, note again that at $T\rightarrow 0$ we recover the usual topological phase $\Phi^0_{\rm U}=\Phi_{\rm B}=\pi$ for $m<c$, and on the other hand, there also exists a critical temperature $T_c$. The FBP corresponds to $m=0$ and $c=1$ where the Majorana modes are completely decoupled from the system dynamics. For the FBP we get the same $T_c$ as before \eqref{Tc} as shown in Fig. \ref{fig_U}(b).

\noindent {\it 5. Polyacetylene (SSH model).}
The following Hamiltonian was introduced in \cite{Rice_Mele82} by Rice and Mele and it has a topological insulating phase:
\begin{equation}
H_{\rm SSH}=-\sum_n\Big (J_1a_n^{\dagger}b_n + J_2a_n^{\dagger}b_{n-1} + \text{H.c.} \Big) + M\sum_n\Big( a_n^{\dagger}a_n-b_n^{\dagger}b_n \Big).
\label{Hpol1}
\end{equation}
The fermionic operators $a_n$ and $b_n$ act on adjacent sites of a dimerized chain. If the energy imbalance between sites $a_n$ and $b_n$ is $M=0$, the above Hamiltonian $H\equiv H_{\rm SSH}$ describes effectively polyacetylene \cite{SSH}, whereas for $M\not=0$ it can model diatomic polymers \cite{Rice_Mele82}.

For $M=0$ and $J_2>J_1$ there are two edge states at the end of the chain and the system displays topological order, characterized by $\omega_1$ and $\Phi_{\rm B}$.

In momentum space, $H_{\rm SSH}$ is written in the form of Eq. \eqref{Hk} with
\begin{eqnarray}
{\bm n}_k&=&\frac{2}{\Delta_k}(-J_1-J_2\cos{k},J_2\sin{k},0),\nonumber\\
\Delta_k&=&2\sqrt{J_1^2+J_2^2+2J_1J_2\cos{k}}.
\label{parMC}
\end{eqnarray}
which in Eq. \eqref{Hk} implies fixing $\theta=\pm\frac{\pi}{2}$ for all $k$.

In Fig. \ref{fig_U}(c) we plot $\Phi_{\rm U}$ as a function of the hopping parameters $J_1$, $J_2$ and the temperature $T$. At $T\rightarrow0$ the topological region coincides again with the usual topological phase, $\Phi^0_{\rm U}=\Phi_{\rm B}=\pi$ for $ J_1<J_2$, and there exists a critical temperature $T_c$.

For the FBP, $J_1=0$ and $J_2=1$, the gap $\Delta_k=2$ becomes constant and we obtain the same critical temperature as for the other two models, Eq. \eqref{Tc}.

\noindent {\it 6. Outlook and Conclusions.}

We have shown that the Uhlmann phase provides us with a way to extend the notion of symmetry-protected topological order in fermion systems beyond the realm of pure states. This comes into play when studying dissipative effects and particularly thermal baths. When applied to three paradigmatic models of topological insulators and superconductors, it displays a discontinuity in some finite critical temperature $T_c$, which limits the region with topological behavior. Interestingly enough, a thermal-bulk-edge correspondence with the Uhlmann phase does not exist, and the topology assessed by it does not determine the fate of the edge modes at finite temperature. 

Although the analysis has been restricted here to 1D models and some representative examples, we expect that the Uhlmann approach could be extended to higher spacial dimensions and other symmetry classes of topological insulators/superconductors. However, more progress on this line is required.

Finally, let us stress that the Uhlmann phase is an observable \cite{Ericsson2003,Sjoqvist_07}. Additionally, we analyze possible experimental measurement schemes in the Supplementary Material \cite{SM}.



\begin{acknowledgments}

We are grateful to the anonymous referee and to Z. Huang for pointing out a correction in a previous version of the manuscript. We thank the Spanish MINECO grant FIS2012-33152, FIS2009-10061, CAM research consortium QUITEMAD S2009-ESP-1594, European Commission
PICC: FP7 2007-2013, Grant No.~249958, UCM-BS grant GICC-910758, FPU MEC Grant and Residencia de Estudiantes.

\end{acknowledgments}


%
%


\newpage

\section*{SUPLEMENTARY MATERIAL}

\appendix

\setcounter{figure}{0}
\setcounter{equation}{0}
\renewcommand*{\thefigure}{S\arabic{figure}}
\renewcommand*{\theequation}{S\arabic{equation}}

\subsection{I.\quad Geometrical Meaning of $T_c$}
\label{app_A}

%

The appearance of a critical temperature $T_c$ in the Uhlmann phase can be better understood from a very simple model for the behavior of the Uhlmann holonomy in fermion systems.

For the sake of simplicity, let us represent the amplitudes (or purifications) as two-dimensional arrows and the phase between two of them as the angle between their corresponding arrows. In Fig. 1, we sketch different behaviors of the amplitudes (arrows) when they are transported according to the Uhlmann's parallel condition along a closed trajectory embracing the whole Brilluoin zone, i.e. from $k=-\pi$ to $k=\pi$, this is left to right on the Fig. 1. We observe several situations:
\begin{itemize}
\item[i)] $T=0$ in the trivial topological regime [Fig. 1(a)]. The arrow is transported with constant slope, the initial and final arrows are parallel, so that $\Phi_{\rm U}^{0}=\Phi_{\rm B}=0$.
\item[ii)] $T=0$ in the non-trivial topological regime [Fig. 1(b)]. The Uhlmann phase remains 0 until some singular point $k_c$, where the direction of the amplitude is suddenly flipped, and so it remains up to the final point $k=\pi$. Thus, the initial arrow and the final arrow form an angle of $\pi$, then $\Phi_{\rm U}^0=\Phi_{\rm B}=\pi$.
\item[iii)]$T_c>T>0$ in a non-trivial topological regime [Fig. 1(c)]. The behavior of the Uhlmann phase follows a similar pattern as before, but now the temperature displaces the position of $k_c$ towards the end of the Brillouin zone.
\item[iv)]$T=T_c$ in a non-trivial topological regime [Fig. 1(d)]. The position of $k_c$ reaches the end of the Brillouin zone.
\item[v)]$T> T _c$ [Fig. 1(a)]. The temperature is so high that a kink never takes place during the trajectory from $k=-\pi$ to $k=\pi$, hence the arrow does not flip and then the Uhlmann phase vanishes $\Phi_{\rm U}=0$.
\end{itemize}

\begin{figure}[t]
\begin{center}
\includegraphics[width=\columnwidth]{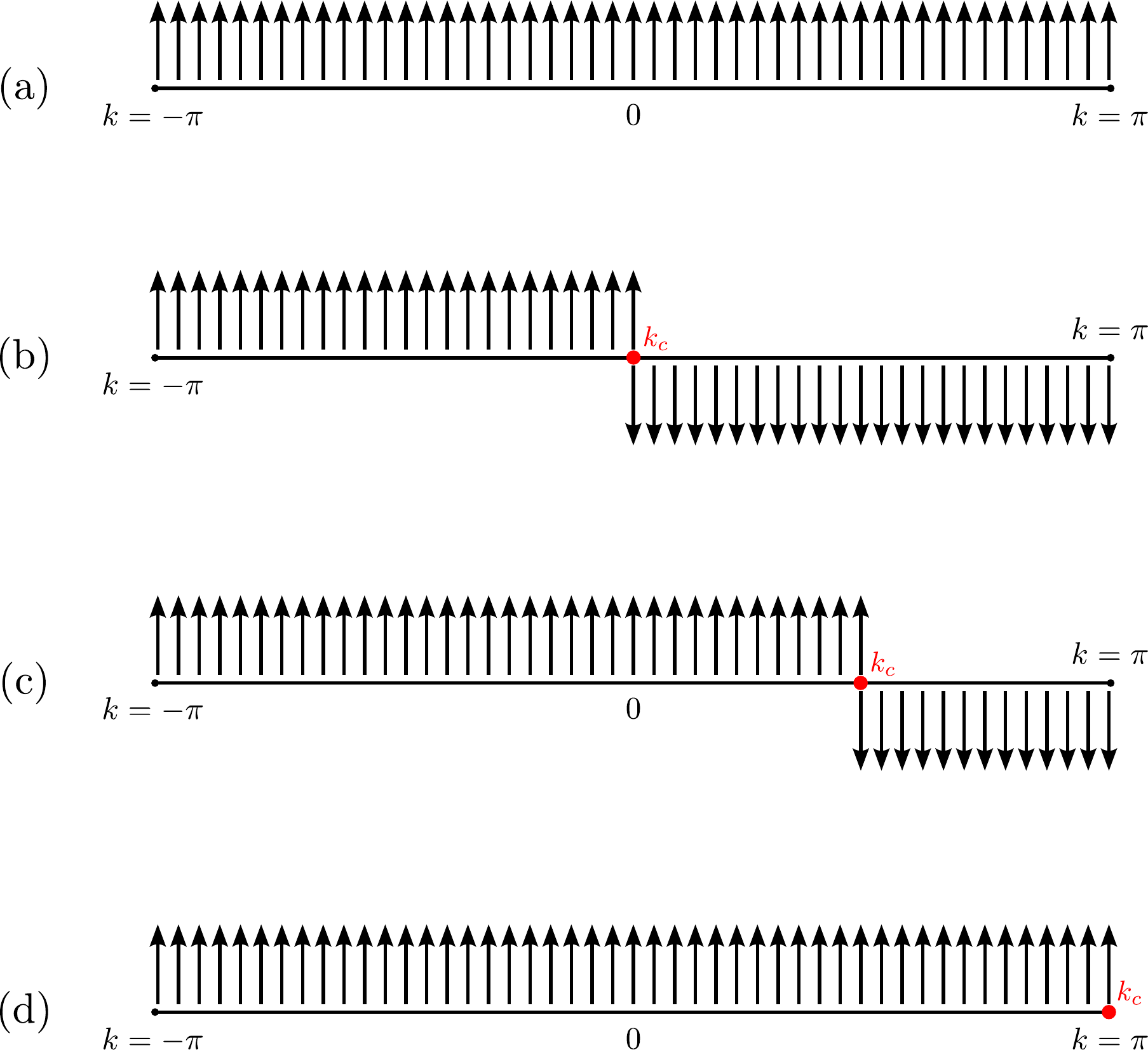}
\end{center}
\caption{Schematic plot for the behavior of the Uhlmann phase during the parallel transport from $k=-\pi$ to $k=\pi$ for different situations: $T=0$ and trivial topological regime or $T>T_c$ (a), $T=0$ and nontrivial topological regime (b), $T_c>T>0$ and nontrivial topological regime (c), and  $T=T_c$ (d). See comments on the main text of this document.}
\end{figure}

Note that the Uhlmann phase places on equal footing $T$ and the parameters of the Hamiltonian $H$. Thus, the position of the critical momentum $k_c$ is affected by both $H$ and $T$.


In summary, as commented in the letter, the existence of a critical temperature $T_c$ is connected to a topological kink structure \cite{Coleman_s} in the Uhlmann holonomy. More precisely, the variation of temperature produces the presence of absence of a topological kink associated to the physical critical momentum $k_c$. In fact, there is not a connection between the Uhlmann phase and the presence or absence of edge states. The critical momentum $k_c$ splits the holonomy into two dis-equivalent topological components according to the value taken by $k$ when performing the trajectory. The first component is the region with $k<k_c$ and $\Phi_{\rm U}=0$, and the second is the region with $k>k_c$ and $\Phi_{\rm U}=\pi$. Note that these two different components cannot be smoothly connected as for that aim the momentum has to cross the singular point $k=k_c$. In the trivial topological regime $\Phi_{\rm U}=0$ for every point along the trajectory and there is only one component. The effect of the temperature in the Uhlmann parallel transport can be understood as a displacement of the topological kink along $k$-space. The critical temperature corresponds to the situation where $k_c$ is at the edge of the Brillouin zone. In other words, $T_c$ determines the admissible amount of noise/disorder such that the Uhlmann holonomy along the Brilluoin zone presents a topological kink structure. For $T>T_c$, the noise is high enough that the kink is effectively ``erased'' and the Uhlmann geometric phase becomes trivial.

\subsection{II.\quad Measurement of the Uhlmann Phase}

As stressed in the letter, the Uhlmann phase is a physical observable. Thus, for the sake of completion we provide here some hints about its measurement in a real experiment. The purpose of this section is to explain how the already existing experimental proposals to measure the Uhlmann phase in the optical domain \cite{Ericsson2003_s,Sjoqvist_07_s} could be adapted to 1D fermionic systems.

Since the Uhlmann holonomy is defined as a relative phase between amplitudes, in order to measure it we need some auxiliary degree of freedom to construct them from density matrices. We hereby present two different ways for this to be implemented.

\subsection{A.\quad Purification approach}

This method is based on the fact that amplitudes $w_k$ can be seen as pure states $\ket{w_k}$ living in an enlarged Hilbert space ${\cal{H}}={\cal{H}}_{\rm S}\otimes{\cal{H}}_{\rm A}$ where both system and ancilla Hilbert spaces are of the same dimension \cite{Ericsson2003_s}.

For instance, we may introduce two electrons in the system with the same crystalline momentum $k_0$ and in the lowest energy band, differing only in a particular non-dynamical degree of freedom. Taking the Creuzt ladder model as an example, the two electrons could differ on their spins $(\uparrow\downarrow)$ and be prepared in their ground states. Once this has been done, we may apply the following steps:
\begin{enumerate}
\item Prepare an entangled state in the diagonal basis of the two electrons 
of the following form:
\begin{align}
\ket{w_{k_0}}&=\sqrt{\tfrac{1+r_{k_0}}{2}}\ket{u^{\uparrow k_0}_{-}}\otimes\ket{u^{\downarrow k_0}_{-}}\nonumber\\
&+\sqrt{\tfrac{1-r_{k_0}}{2}}\ket{u^{\uparrow k_0}_{+}}\otimes\ket{u^{\downarrow k_0}_{+}},
\end{align}
where $r_{k_0}:=\tanh{\beta\frac{\Delta_{k_0}}{2}}$. Let us take for simplicity the flat-band case where $\Delta_k=2$. Therefore, once the temperature is fixed, $r_k$ is a constant during the holonomy. By taking partial trace with respect to the ancillary electron, the reduced state for the system electron is
\begin{equation}
\rho_{k_0}=\Tr_{\rm A} (\ket{w_{k_0}}\bra{w_{k_0}})=\frac{1}{2}\Big(\mathds{1}-r_{k_0}\sigma_z\Big).
\end{equation}
This corresponds to a thermal state of the system written in the diagonal basis of the system Hamiltonian.
\item Implement the holonomy in $k-$space. For example by applying the unitary operation $V_{k(t)}$ on the ancillary electron. This is determined by imposing the Uhlmann's parallel transport condition on the trajectory $\{k(t)\}_{t=0}^1$ such that $k(0)=k_0$ and $k(1)=k_0+G$, where $G$ stands for the reciprocal lattice vector. This might be achieved using a spin dependent force.
\item Interferometry to measure the relative phase. To retrieve the Uhlmann phase, we make use of
\begin{equation}
\Phi_{\rm U}=\arg \langle w_{k(0)}|w_{k(1)}\rangle.
\end{equation}
This can be implemented using atom interferometry techniques similar to those in \cite{Atala_et_al_12_s}. Another example where the degree of control over fermionic systems is at the highest level can be found in \cite{Tarruell_et_al12_s}.
\end{enumerate}

\subsection{B.\quad System plus ancillary qubit}

A different approach to construct amplitudes was proposed for the optical domain in \cite{Sjoqvist_07_s}. The idea is again to enlarge the Hilbert space ${\cal{H}}={\cal{H}}_{\rm S}\otimes{\cal{H}}_{{\rm qubit}}$. However, instead of preparing two copies of the system, it is just required an auxiliary two-dimensional quantum system (qubit). Then system and qubit are prepared in a certain mixed state $\hat{\rho}$. It can be shown that the amplitudes associated to the state $\rho_k=w_kw^{\dagger}_k$ appear in the coherences of this larger state $\hat{\rho}_k$ \cite{Sjoqvist_07_s}:
\begin{align}
\hat{\rho}_k&=\frac{1}{2}\rho_k\otimes\ket{0}\bra{0}+ \frac{1}{2n}\mathds{1}\otimes\ket{1}\bra{1}\nonumber\\
&+\frac{1}{2\sqrt{n}}w_k\otimes\ket{0}\bra{1}+\frac{1}{2\sqrt{n}}w^{\dagger}_k\otimes\ket{1}\bra{0},
\end{align}
where $n$ is the dimension of the system, i.e. the number of bands for the fermion models. Secondly, a protocol can be designed \cite{Sjoqvist_07_s} in order to implement the unitary $V_{k(t)}$ on the amplitude and retrieve the Uhlmann phase after the holonomy using again atom interferometric techniques.

Therefore, we can envision possible measurement schemes for the Uhlmann phase in fermionic systems based on what has already been proposed previously for photons in the context of geometric phases for qubits. However, giving a precise experimental proposal for a particular setup would require further analysis which is beyond the scope of this paper.

\end{document}